\documentclass[12pt,a4paper]{article}
\usepackage{amssymb,amsmath}
\usepackage[dvips]{lscape,graphicx}

\newcommand{\bc}{\begin{center}}
\newcommand{\ec}{\end{center}}
\newcommand{\bd}{\begin{displaymath}}
\newcommand{\ed}{\end{displaymath}}
\newcommand{\be}{\begin{equation}}
\newcommand{\ee}{\end{equation}}
\newcommand{\ba}{\begin{array}}
\newcommand{\ea}{\end{array}}
\newcommand{\bt}{\begin{tabular}}
\newcommand{\et}{\end{tabular}}

\newcommand{\ds}{\displaystyle}

\begin{document}

\title{On the smallness of the cosmological constant in SUGRA models}

\author{C.Froggatt${}^{1}$, R.Nevzorov${}^{2}$, H.B.Nielsen${}^{3}$\\[5mm]
\itshape{${}^{1}$ Department of Physics and Astronomy, Glasgow University, Scotland}\\[0mm]
\itshape{${}^{2}$ School of Physics and Astronomy, University of Southampton, UK}\\[0mm]
\itshape{${}^{3}$ The Niels Bohr Institute, Copenhagen, Denmark}}

\date{}

\maketitle

\begin{abstract}{
\noindent In no--scale supergravity global symmetries protect
local supersymmetry and a zero value for the cosmological
constant. We consider the breakdown of these symmetries and
present a minimal SUGRA model motivated by the multiple point
principle, in which the total vacuum energy density is naturally
tiny. In order to reproduce the observed value of the cosmological
constant and preserve gauge coupling unification, an additional
pair of $5+\bar{5}$--plets of superfields has to be included in
the particle content of the considered model. These extra fields
have masses of the order of the supersymmetry breaking scale; so
they can be detected at future colliders. We also discuss the
supersymmetry breakdown and possible solution of the cosmological
constant problem by MPP in models with an enlarged gauge
symmetry.}
\end{abstract}

\vspace{6.5cm} \footnoterule{\noindent
${}^{1}$ E-mail: c.froggatt@physics.gla.ac.uk\\
${}^{2}$ On leave of absence from the Theory Department, ITEP, Moscow, Russia;\\
$~~$ E-mail: nevzorov@phys.soton.ac.uk\\
${}^{3}$ E-mail: hbech@alf.nbi.dk}

\section{Introduction}

The origin of a tiny energy density spread all over the Universe
(the cosmological constant $\Lambda$), which is responsible for
its accelerated expansion, is one of the most challenging problems
nowadays. A fit to the recent data shows that $\Lambda \sim
10^{-123}M_{Pl}^4 \sim 10^{-55} M_Z^4$ \cite{1}. At the same time
the presence of a gluon condensate in the vacuum is expected to
contribute an energy density of order $\Lambda_{QCD}^4\simeq
10^{-74}M_{Pl}^4$. On the other hand if we believe in the Standard
Model (SM) then a much larger contribution $\sim v^4\simeq
10^{-62}M_{Pl}^4$ must come from the electroweak symmetry
breaking. The contribution of zero--modes is expected to push the
vacuum energy density even higher up to $\sim M_{Pl}^4$. Therefore
the smallness of the cosmological constant should be regarded as a
fine-tuning problem, for which new theoretical ideas must be
employed to explain the enormous cancellations between the
contributions of different condensates to the total vacuum energy
density.

At this moment none of the available generalizations of the SM
provides a satisfactory explanation for the smallness of the
cosmological constant. An exact global supersymmetry (SUSY)
ensures zero value for the energy density at the minimum of the
potential of the scalar fields. However, in the exact SUSY limit,
bosons and fermions from one chiral multiplet are degenerate.
Because superpartners of quarks and leptons have not been observed
yet, supersymmetry must be broken. In general the breakdown of
supersymmetry induces a huge and positive contribution to the
total vacuum energy density of order $M_{S}^4$, where $M_{S}$ is
the SUSY breaking scale. The non--observation of superpartners of
observable fermions implies that $M_{S}\gg 100\,\mbox{GeV}$.

Our basic scenario for evaluating the tiny value of the
cosmological constant is to assume the existence of a second
vacuum degenerate with the one in which we live. We assume that
our vacuum is a softly broken supersymmetric vacuum and that the
second vacuum is supersymmetric without any soft SUSY breaking
terms. However we imagine that the supersymmetry in the second
vacuum is broken dynamically, when the supersymmetric QCD
interaction becomes non-perturbative. This happens at a much lower
energy scale than $\Lambda_{QCD}$, since the supersymmetric QCD
beta function must be used, and thereby generates a small
cosmological constant. This small value is then transferred to our
vacuum by the assumed degeneracy.

The assumed degeneracy of the vacua is supposed to be justified by
the so-called Multiple Point Principle (MPP) \cite{10}, according
to which Nature chooses values of coupling constants such that
many phases of the underlying theory should coexist. On the phase
diagram of the theory it corresponds to the special point -- the
multiple point -- where many phases meet. The vacuum energy
densities of these different phases are degenerate at the multiple
point.

In the case of global supersymmetry, the energy density of a
supersymmetric vacuum is naturally zero. However, since we are
interested in the value of the cosmological constant, we must
include gravity and thus local supersymmetry. In supergravity
(SUGRA) models, the vacuum energy density is not naturally zero;
indeed in general it is expected to be large and negative. In our
MPP scenario above, prior to the dynamical SUSY breaking in the
second vacuum, we require the existence of degenerate
supersymmetric and non-supersymmetric vacua with vanishing energy
density. In a previous application of MPP to supergravity
\cite{18}, a supersymmetric phase in flat Minkowski space was
simply assumed to exist, in addition to the phase in which we
live. Since the vacuum energy density of supersymmetric states in
flat Minkowski space is just zero, the cosmological constant
problem was thereby solved to first approximation by assumption.
The degeneracy between the supersymmetric and physical vacua was
attained by fine-tuning the K$\Ddot{a}$hler function of the
considered SUGRA model \cite{18}. However this previous work
corresponds to searching for only a partial solution of the
cosmological constant problem and makes the whole approach look
rather artificial. The situation changes dramatically if
supergravity can be supplemented by a global symmetry that ensures
a zero value for the cosmological constant. This is precisely what
happens in no-scale supergravity.

In no-scale supergravity the supersymmetric states with zero
vacuum energy density emerge automatically at low energies. But
the global symmetry, which ensures the vanishing of the
cosmological constant and the degeneracy of global vacua in the
no-scale models, also protects supersymmetry which has to be
broken in any phenomenologically acceptable theory. In this paper
we explore no--scale SUGRA models in which the extended global
symmetry is broken in such a way that our MPP scenario is
fulfilled without any extra fine-tuning. In the next section we
specify the no--scale SUGRA models, consider the breakdown of
local supersymmetry in these models and discuss the connection
with MPP.

The simplest model, in which the implementation of our MPP
scenario does not require any extra fine-tuning, is constructed in
section \ref{minimal}. In section \ref{cosmological} we estimate
the value of the cosmological constant in
MPP inspired SUGRA models. The
realization of our MPP scenario in models based on enlarged gauge
symmetry groups like $\biggl[SU(3)\times SU(2)\times
U(1)\biggr]^3$ is considered in section \ref{extended}. Section
\ref{conclusion} is reserved for our conclusions and outlook.

\section{No--scale supergravity}
\label{noscale}

The full $(N=1)$ SUGRA Lagrangian \cite{4}-\cite{3} is specified
in terms of an analytic gauge kinetic function $f_a(\phi_{M})$ and
a real gauge-invariant K$\Ddot{a}$hler function
$G(\phi_{M},\phi_{M}^{*})$, which depend on the chiral superfields
$\phi_M$. The function $f_{a}(\phi_M)$ determines the kinetic
terms for the fields in the vector supermultiplets and the gauge
coupling constants $Re f_a(\phi_M)=1/g_a^2$, where the index $a$
designates different gauge groups. The K$\Ddot{a}$hler function is
a combination of two functions
\be
G(\phi_{M},\phi_{M}^{*})=K(\phi_{M},\phi_{M}^{*})+\ln|W(\phi_M)|^2\,,
\label{1}
\ee
where $K(\phi_{M},\phi_{M}^{*})$ is the
K$\Ddot{a}$hler potential whose second derivatives define the
kinetic terms for the fields in the chiral supermultiplets.
$W(\phi_M)$ is the complete analytic superpotential of the
considered SUSY model. Here we shall use standard supergravity
mass units: $\ds\frac{M_{Pl}}{\sqrt{8\pi}}=1$.

The SUGRA scalar potential can be presented as a sum of $F$-- and
D--terms
$V_{SUGRA}(\phi_M, \phi^{*}_M)=V_{F}(\phi_M, \phi^{*}_M)+V_{D}(\phi_M, \phi^{*}_M)$,
where the F-- and D--parts are given by \cite{3}-\cite{5}
\be
\ba{rcl}
V_{F}(\phi_M,\phi^{*}_M)&=&\sum_{M,\,\bar{N}} e^{G}\left(G_{M}G^{M\bar{N}}
G_{\bar{N}}-3\right)\, ,\\[3mm]
V_{D}(\phi_M,\phi^{*}_M)&=&\ds\frac{1}{2}\sum_{a}(D^{a})^2\,,
\qquad D^{a}=g_{a}\sum_{i,\,j}\left(G_i
T^a_{ij}\phi_j\right)\\[3mm]
G_M \equiv\partial_{M} G&\equiv&\partial G/\partial \phi_M,
\qquad G_{\bar{M}}\equiv
\partial_{\bar{M}}G\equiv\partial G/ \partial \phi^{*}_M\, .
\ea
\label{2}
\ee
In Eq.~(\ref{2}) $g_a$ is the gauge coupling constant associated with
the generator $T^a$ of the gauge transformations.
The matrix $G^{M\bar{N}}$ is the inverse of the K$\Ddot{a}$hler
metric $K_{\bar{N}M}$, i.e.
$$
G_{\bar{N}M}\equiv\partial_{\bar{N}}\partial_{M}G=
\partial_{\bar{N}}\partial_{M}K\equiv
K_{\bar{N}M}\,.
$$

In order to break supersymmetry in $(N=1)$ SUGRA models,
a hidden sector is introduced.
It contains superfields $(z_i)$, which are singlets under the SM
$SU(3)_C\times SU(2)_W\times U(1)_Y$ gauge group.
It is assumed that the superfields of the hidden sector interact
with the observable ones only by means of gravity.
If, at the minimum of the scalar potential, hidden sector fields
acquire vacuum expectation values so that
at least one of their auxiliary fields
\be
F^{M}=e^{G/2}G^{M\bar{P}}G_{\bar{P}}
\label{3}
\ee
is non-vanishing, then local SUSY is spontaneously broken. At the same time a
massless fermion with spin $1/2$ -- the goldstino, which is a combination of
the fermionic partners of the hidden sector fields giving rise to the breaking
of SUGRA, is swallowed up by the gravitino which thereby becomes massive
$m_{3/2}=<e^{G/2}>$. This phenomenon is called the super-Higgs effect \cite{6}.

Usually the vacuum energy density at the minimum of SUGRA scalar potential
(\ref{2}) is negative. To show this, let us suppose that, the
K$\ddot{a}$hler function has a stationary point, where all derivatives
$G_M=0$. Then it is easy to check that this point is
also an extremum of the SUGRA scalar potential. In the vicinity of
this point local supersymmetry remains intact while the energy
density is $-3<e^{G}>$. It implies that the vacuum energy density must
be less than or equal to this value. Therefore, in general, an enormous
fine--tuning must be imposed, in order to keep the total vacuum energy
density in SUGRA models around the observed value of the cosmological
constant \cite{2}.

Because the smallness of the parameters in a physical theory may be related
to an almost exact symmetry, it is interesting to
investigate what kind of symmetries could protect the cosmological constant
in $N=1$ supergravity. It was discovered
a long time ago that invariance with respect to $SU(1, 1)$ symmetry
transformations results in a tree--level
scalar potential which vanishes identically along some directions
\cite{4}, \cite{11}-\cite{12}. In other words the corresponding
scalar potential (\ref{2}) possesses an infinite set of degenerate
minima with zero vacuum energy density. The $SU(1,1)$
structure of the $N=1$ SUGRA Lagrangian can have its roots in supergravity
theories with extended supersymmetry ($N=4$ or $N=8$) \cite{4}.

The group $SU(1,1)$ contains subgroups of imaginary translations and
dilatations \cite{12}--\cite{15}. The invariance of the K$\ddot{a}$hler
function under the imaginary translations of the hidden sector superfields
\be
z_i\to z_i+i\beta_i\,;\qquad
\varphi_{\alpha}\to\varphi_{\alpha}
\label{4}
\ee
implies that the K$\ddot{a}$hler potential depends only on $z_i+\bar{z}_i$,
while the superpotential is given by \cite{16}
\be
W(z_i,\varphi_{\alpha})=\exp\left\{\sum_{i=1}^{m} a_i z_i\right\}\tilde{W}(\varphi_{\alpha})\,,
\label{5}
\ee
where the $a_i$ are real. Here we assume that the hidden sector
involves $m$ singlet superfields while the observable sector
comprises chiral multiplets $\varphi_{\alpha}$. Since
$G(\phi_M,\bar{\phi}_M)$ is evidently invariant under the
K$\ddot{a}$hler transformations \cite{17}
$$
\left\{
\ba{l}
K(\phi_M,\bar{\phi}_M)\to K(\phi_M,\bar{\phi}_M)-g(\phi_M)
-g^{*}(\bar{\phi}_M)\,,\\[2mm]
W(\phi_M)\to \ds e^{g(\phi_M)}W(\phi_M)
\ea
\right.\,.
$$
the most general K$\ddot{a}$hler function can be written as
\be
G(\phi_M,\bar{\phi}_M)=K(z_i+\bar{z}_i,\varphi_{\alpha},
\bar{\varphi}_{\alpha})+\ln|W(\varphi_{\alpha})|\,,
\label{6}
\ee
where $W(\varphi_{\alpha})=\tilde{W}(\varphi_{\alpha})$.

The dilatation invariance constrains the K$\ddot{a}$hler potential
and superpotential further. Suppose that hidden and observable
superfields transform differently 
\be 
z_i\to\alpha^2 z_i\,,\qquad
\varphi_{\sigma}\to\alpha\varphi_{\sigma}\,. 
\label{7} 
\ee 
Then the structure of the superpotential $W(\varphi_{\alpha})$ in
phenomenologically acceptable SUGRA models is determined by the
symmetry transformations (\ref{4}) and (\ref{7}). Indeed because
the superpotential in these models contains trilinear terms, which
induce masses of quarks and leptons, all terms involving $n$
chiral superfields with $n\gtrless 3$ are forbidden by the
dilatation invariance. If there is only one field $T$ in the
hidden sector, then the K$\ddot{a}$hler function is fixed uniquely
by the gauge and global symmetries of the model:
\be 
\ba{c}
K(T+\bar{T},\varphi_{\sigma},\bar{\varphi}_{\sigma})=\ds-3 \ln(T+\bar{T})+
\sum_{\sigma} C_{\sigma}\frac{|\varphi_{\sigma}|^2}{(T+\bar{T})}\\[2mm]
W(\varphi_{\alpha})=\ds\sum_{\sigma,\beta,\gamma}\ds\frac{1}{6}
Y_{\sigma\beta\gamma}\varphi_{\sigma}\varphi_{\beta}\varphi_{\gamma}\,,
\ea 
\label{8} 
\ee 
where $C_{\sigma}$ and $Y_{\sigma\beta\gamma}$
are constants. Here we restrict our consideration to the lowest
order terms $|\varphi_{\sigma}|^2$ in the expansion of the
K$\ddot{a}$hler potential in terms of observable superfields. The
contribution of higher order terms to the SUGRA scalar potential
is suppressed by inverse powers of $M_{Pl}$ and can be safely
ignored.

For the particular choice of the symmetry transformations (\ref{7})
the part of the SUGRA scalar potential which is induced by the 
K$\ddot{a}$hler function of the hidden sector vanishes \cite{12}, i.e.
$$
V_{hid}=\ds e^{G}\left(G_{T}G^{T\bar{T}}G_{\bar{T}}-3\right)=0\,.
$$
Then the full scalar potential takes the form
\be
V=\frac{1}{3}e^{2K/3}\sum_{\alpha}\biggl|\ds\frac{\partial
W(\tilde{\varphi}_{\alpha})}{\partial\tilde{\varphi}_{\alpha}}
\biggr|^2+\ds\frac{1}{2}\sum_{a}(D^{a})^2\,,
\label{9}
\ee
where the observable superfields are rescaled as
$\tilde{\varphi}_{\alpha}=\ds\sqrt{\frac{C_{\sigma}}{3}}\varphi_{\alpha}$.
The potential (\ref{9}) leads to a supersymmetric particle
spectrum at low energies. Owing to the particular form of the
K$\ddot{a}$hler potential (\ref{8}) with $k=2$, it is positive definite.
Its minimum is reached at the points for which
$\biggl<\ds\frac{\partial W(\varphi_{\alpha})}
{\partial \varphi_{\alpha}}\biggr>=\,<D^{a}>=0$.
As a consequence the vacuum energy density goes to zero near global
minima of the scalar potential (\ref{9}). Thus
imaginary translations (\ref{4}) and dilatations (\ref{7})
protect a zero value for the cosmological constant 
in supergravity \cite{12} \footnote{In \cite{91} a symmetry that forbids a
cosmological constant in six and ten dimensional theories is discussed.}.

The invariance of the K$\ddot{a}$hler function with respect to
symmetry transformations (\ref{4}) and (\ref{7}) also prevents
the breaking of local supersymmetry. In order to illustrate this,
let us consider an SU(5) SUSY model with one field in
the adjoint representation $\Phi$ and with one singlet field $S$.
As before the structure of the K$\ddot{a}$hler function is
completely fixed by the global symmetries (\ref{4}) and (\ref{7}),
which result in a K$\Ddot{a}$hler potential and superpotential of
the form given by Eq.~(\ref{8}). The superpotential of
the considered model is further constrained by the
$SU(5)$ gauge symmetry:
\be
W(S,\Phi)=\ds\frac{\varkappa}{3}S^3+\lambda \mbox{Tr}\Phi^3+\sigma
S \mbox{Tr}\Phi^2\,.
\label{10}
\ee
In the general case the minimum of the scalar potential,
which is induced by the superpotential (\ref{10}), is
attained when $<S>=<\Phi>=0$ and does not lead to the
breakdown of local supersymmetry or of gauge symmetry. But if
$\varkappa=-40\sigma^3/(3\lambda^2)$ there is a vacuum configuration
\be
<\Phi>=\ds\frac{\Phi_0}{\sqrt{15}}\left(
\ba{ccccc}
1 & 0 & 0 & 0 & 0 \\[0mm]
0 & 1 & 0 & 0 & 0 \\[0mm]
0 & 0 & 1 & 0 & 0 \\[0mm]
0 & 0 & 0 & -3/2 & 0 \\[0mm]
0 & 0 & 0 & 0 & -3/2
\ea
\right)\,,\qquad
\ba{l}
<S>=S_0\,, \\
\\
\Phi_0=\ds\frac{4\sqrt{15}\sigma}{3\lambda}S_0\,,
\ea
\label{11}
\ee
which breaks SU(5) down to $SU(3)\times SU(2)\times U(1)$.
However, along the valley (\ref{11}), the superpotential and all
auxiliary fields $F_i$ vanish preserving local supersymmetry and
the zero value of the vacuum energy density.

In order to get a vacuum where local supersymmetry is broken, one
should violate dilatation invariance in the superpotential.
Eliminating the singlet field from the considered SU(5) model and
introducing a mass term for the adjoint representation, we get
the superpotential
\be
W(\Phi)=M_X\mbox{Tr}\Phi^2+\lambda \mbox{Tr}\Phi^3\,.
\label{12}
\ee
The scalar potential of the resulting model is given by Eq.~(\ref{9}).
It has a few degenerate vacua with vanishing vacuum energy density.
For example, in the scalar potential there exist a minimum where
$<\Phi>=0$ and another vacuum, which has a configuration similar to
Eq.~(\ref{11}) but with $\Phi_0=\ds\frac{4\sqrt{15}}{3\lambda}M_X$.
In the first vacuum the SU(5) symmetry and local supersymmetry
remain intact, while in the second one the auxiliary field
$F_{T}$ acquires a vacuum expectation value and a non-zero
gravitino mass is generated:
\be
\ba{rclcl}
<|F_T|>&\simeq&\left<\ds\frac{|W(\Phi)|}{(T+\bar{T})^{1/2}}\right>&
=&m_{3/2}\left<(T+\bar{T})\right>\,,\\[3mm]
m_{3/2}&=&\left<\ds\frac{|W(\Phi)|}{(T+\bar{T})^{3/2}}\right>&
=&\ds\frac{40}{9}\frac{M_X^3}{\lambda^2\left<(T+\bar{T})^{3/2}\right>}\,
\ea
\label{13}
\ee
although the vacuum expectation value of $T$ is undetermined at
tree level, since the hidden sector scalar potential is flat.
As a result, local supersymmetry and gauge symmetry are broken in
the second vacuum. Nevertheless the invariance of the low energy effective
Lagrangian of the observable sector under the transformations of
global supersymmetry is preserved (see Eq.~(\ref{9})). When $M_X$
goes to zero the dilatation invariance, as well as SU(5) symmetry
and local supersymmetry in the second vacuum, are restored.

This simple SU(5) model with the superpotential (\ref{12})
illustrates how the degenerate vacua required for the application
of MPP to supergravity are naturally realized in no-scale
supergravity. In the second vacuum local supersymmetry is broken,
as is supposed to be the case in the physical vacuum in which we
live. It is usually supposed that local supersymmetry breaking
induces SUSY breaking terms. However there are no such terms in
this no-scale SUGRA model and global supersymmetry is unbroken in
both vacua.

\section{Minimal MPP inspired SUGRA model}
\label{minimal}

The no-scale SUGRA model with the superpotential (\ref{12}) is not viable
from the phenomenological point of view, due to the absence of global
supersymmetry breaking in the observable sector for all vacua.
This raises the question of whether it is possible to construct
a phenomenologically acceptable model based on broken
global symmetries (\ref{4}) and (\ref{7})
, which realises our MPP scenario without
extra fine--tuning. We need to generate soft SUSY breaking terms
that break global supersymmetry in the observable sector of the
physical vacuum. These soft terms are generally characterised by
the gravitino mass scale, which must then be of order the
electroweak scale. This required small value of the gravitino mass
$m_{3/2}$ of course constitutes the gauge hierarchy problem, whose
solution was the original motivation for no-scale models with a
flat hidden sector scalar potential\footnote{
An enormous mass hierarchy ($m_{3/2}\ll M_{Pl}$) can appear due to
a non-perturbative source of local supersymmetry breaking \cite{9}.}.
In this paper we concentrate
on the hierarchy problem associated with the tiny value of the
cosmological constant and do not explicitly address the solution
of the gauge hierarchy problem. We shall simply assume there is a
weak breaking of the dilatation invariance of the hidden sector
superpotential characterised by an hierarchically small parameter
$\varkappa$.

In fact, we take the hidden sector to include two superfields,
$T$ and $z$, that transform differently under dilatations
\be
T\to \alpha^2 T\,,\qquad z\to \alpha z\,,\qquad
\varphi_{\alpha}\to \alpha \varphi_{\alpha}
\label{18}
\ee
and imaginary translations
\be
T\to T+i\beta\,,\qquad z\to z\,,\qquad
\varphi_{\alpha}\to\varphi_{\alpha}\,.
\label{19}
\ee
In Eq.~(\ref{18})--(\ref{19}) $\varphi_{\alpha}$ represent
the observable superfields. The hidden sector superfield $z$
transforms similarly to $\varphi_{\alpha}$ under the global symmetry
transformations (\ref{18})--(\ref{19}). It plays a role analogous
to the SU(5) adjoint field $\Phi$ in Eq.~(\ref{12}) and appears
in the full superpotential of the model:
\be
W(z,\,\varphi_{\alpha})=\varkappa\biggl(z^3+
\mu_0 z^2+\sum_{n=4}^{\infty}c_n z^n\biggr)+
\ds\sum_{\sigma,\beta,\gamma}\ds\frac{1}{6}
Y_{\sigma\beta\gamma}\varphi_{\sigma}\varphi_{\beta}\varphi_{\gamma}\,,
\label{191}
\ee

The bilinear mass term for the superfield $z$ and the higher
order terms $c_n z^n$ in the superpotential (\ref{191})
spoil the dilatation invariance. But, as we noticed in
section \ref{noscale}, such a breakdown of the symmetry
protecting the cosmological constant may preserve a zero
value of the vacuum energy density in all global minima
of the scalar potential of the model, if the structure
of the K$\Ddot{a}$hler potential remains intact. It
may also give rise to the spontaneous breakdown of
local supersymmetry in the physical vacuum.
Furthermore we require a locally supersymmetric
vacuum with zero cosmological constant in our MPP scenario.
We note that the conditions for the existence of such a
vacuum are that the superpotential $W$ for the hidden sector
and its derivatives should vanish\footnote{The vanishing of
$W$ implies that the last term in the expression for
$V_{F}(\phi_M,\phi^{*}_M)$ (see Eq.~(\ref{2})), which led
to the negative energy density, vanishes. Taking into account
that the K$\Ddot{a}$hler metric of the hidden sector is
positive definite, one can prove that the absolute minimum
of the scalar potential (\ref{2}) is achieved when the derivative
of $W$ vanishes \cite{18}.} at the corresponding minimum of the scalar
potential:
\be
\biggl< W (z)\biggr>=\biggl<\frac{\partial W(z)}{\partial z}\biggr>=0\,.
\label{15}
\ee
So we restrict our considerations to breakdowns of dilatation
invariance which result in a global minimum of the SUGRA scalar
potential at $z=0$, because it represents a vacuum where
local supersymmetry remains intact. According to Eq.~(\ref{9})
there is no  global minimum at $z=0$,
if the superpotential (\ref{191}) contains a term
proportional to $z$ or terms which are inversely proportional
to a power of $z$. Terms involving negative powers of the
superfields are not present in the superpotentials of the
simplest SUSY models like the minimal supersymmetric standard
model (MSSM) and the next to minimal supersymmetric standard
model. A term proportional to $z$ can be forbidden by a gauge
symmetry of the hidden sector, if $z$ transforms
non--trivially under the corresponding gauge transformations,
as in the case of our toy $SU(5)$ model (\ref{12}).

Because the dilatation invariance is broken explicitly,
one may expect the appearance of bilinear and higher order
terms in the superpotential of the observable sector.
Some of them are potentially dangerous. For instance,
the inclusion of the bilinear terms
$\mu_{\alpha\beta}\varphi_{\alpha}\varphi_{\beta}$
leads to the so--called $\mu$--problem in the simplest
SUSY models. Actually in the MSSM, the SM
gauge symmetry allows only one bilinear term
$\mu H_1 \epsilon H_2$ where $H_1$ and $H_2$ are Higgs
doublets. From dimensional considerations it is obvious
that the corresponding mass parameter $\mu$ should be
of order of the Planck scale, because this is the only
scale characterising SUGRA theories. At the same time
the correct pattern of electroweak symmetry breaking
requires $\mu$ to be in the $\mbox{TeV range}$.
In order to avoid a ``new hierarchy'' problem,
the dilatation invariance should not be spoilt in the
part of the superpotential
(\ref{191}) that includes observable superfields.

For completness we have to specify the K$\Ddot{a}$hler
potential in our MPP inspired SUGRA model.
It is fixed as follows
\be
\ba{rcl}
K(\phi_{M},\phi_{M}^{*})&=&\ds-3\ln\biggl[T+\overline{T}
-|z|^2-\sum_{\alpha}\zeta_{\alpha}|\varphi_{\alpha}|^2\biggr]
+\\[2mm]
&+&\sum_{\alpha, \beta}\biggl(\ds\frac{\eta_{\alpha\beta}}
{2}\,\varphi_{\alpha}\,\varphi_{\beta}+h.c.\biggr)+
\sum_{\beta}\xi_{\beta}|\varphi_{\beta}|^2\,,
\ea
\label{192}
\ee
where $\zeta_{\alpha}$, $\eta_{\alpha\beta}$, $\xi_{\beta}$
are some constants. The kinetic terms of the scalar fields,
which are induced by the first term on the right hand side
of Eq.~(\ref{192}), are invariant under the isometric
transformations of the non--compact $SU(N,1)$ group \cite{13},
where $N$ is the number of chiral superfields in the model.
This symmetry can be derived from extended ($N\ge 5$)
supergravity theories \cite{14}. The Yukawa interactions
in the superpotential (\ref{191}) and D--terms in the scalar
potential break $SU(N,1)$ symmetry explicitly, in such a
way that only invariance under the dilatations and imaginary
translations can be realized in phenomenologically
viable $N=1$ SUGRA models. Exactly this type of SUGRA model
was discussed in section \ref{noscale}. The K$\Ddot{a}$hler
potential (\ref{8}) can be easily reproduced,
if one expands the first term in Eq.~(\ref{192}) in powers
of $\ds\frac{|z|^2}{T+\overline{T}}$ and
$\ds\frac{|\varphi_{\alpha}|^2}{T+\overline{T}}$.
Thus, in the limit when $\eta_{\alpha\beta}$, $\xi_{\beta}$
and $\varkappa$ go to zero, the invariance under the
symmetry transformations (\ref{18})--(\ref{19}) is
restored, protecting supersymmetry and a zero value
of the cosmological constant.

In section \ref{noscale} we demonstrated that the violation of
dilatation invariance does not necessarily cause the breaking of
global supersymmetry at low energies. This is the reason why we
include extra terms in the K$\Ddot{a}$hler potential of our SUGRA
model. We allow the breakdown of the dilatation invariance in the
K$\Ddot{a}$hler potential of the observable sector only. The part
of $K(\phi_{M},\phi_{M}^{*})$ involving hidden sector superfields
is responsible for the cancellation of the negative contribution
to the total vacuum energy density coming from the term $-3e^{G}$
in the scalar potential (\ref{2}). Therefore any variations in the
K$\Ddot{a}$hler potential of the hidden sector may spoil the
vanishing of the vacuum energy density in global minima. 
For example, if the factor in front of the logarithm in 
Eq.~(\ref{192}) is greater than $-3$ then SUGRA scalar potential 
is not positive definite and the total energy density tends to be
huge and negative.

In order to avoid cumbersome calculations, we introduce
the simplest set of terms breaking the dilatation invariance
in the K$\Ddot{a}$hler potential. All the terms are
bilinear with respect to observable superfields and
do not depend on the hidden sector fields. Higher
order terms are irrelevant for our study, since
their contribution to the low energy effective
potential is suppressed by inverse powers of $M_{Pl}$.
Additional terms which are proportional to
$|\varphi_{\alpha}|^2$ normally appear in minimal
SUGRA models \cite{21}--\cite{22}. The other terms
$\eta_{\alpha\beta}\varphi_{\alpha} \varphi_{\beta}$
introduced in the K$\Ddot{a}$hler potential (\ref{192})
give rise to effective $\mu$ terms after the spontaneous
breakdown of local supersymmetry, solving the $\mu$ problem \cite{30}.

In the limit when $\xi_{\beta}$ and $\eta_{\alpha\beta}$
vanish while $\zeta_{\alpha}\to 1$, we return back to the
SUGRA scalar potential of the form (\ref{9}). In this case,
the scalar potential of the hidden sector becomes
\be
V(T,\, z)=\frac{1}{3(T+\overline{T}-|z|^2)^2}
\biggl|\frac{\partial W(z)}{\partial z}\biggr|^2\,.
\label{193}
\ee
The minima of the scalar potential (\ref{193}) are attained
at the stationary points of the hidden sector superpotential.
In the simplest case when $c_n=0$, the superpotential
(\ref{191}) has two extremum points at $z=0$ and
$z=-\ds\frac{2\mu_0}{3}$. At these points the scalar
potential (\ref{193}) achieves its absolute minimal value
i.e.~zero. In the first vacuum where $z=-\ds\frac{2\mu_0}{3}$,
local supersymmetry is broken and the gravitino gets a
non--zero mass:
\be
m_{3/2}=\biggl<\frac{W(z)}{(T+\overline{T}-|z|^2)^{3/2}}\biggr>
=\frac{4\varkappa\mu_0^3}{27\biggl<\biggl(T+\overline{T}
-\ds\frac{4\mu_0^2}{9}\biggr)^{3/2}\biggr>}\,.
\label{194}
\ee
In the second minimum, the vacuum expectation value of the
superfield $z$ and the superpotential of the hidden sector vanish.
Therefore the conditions (\ref{15}) are fulfilled automatically
and local supersymmetry remains intact. If the high order terms
$c_n z^n$ are present in Eq.~(\ref{191}), the scalar potential
of the hidden sector may have many degenerate vacua with
vanishing vacuum energy density, where the gravitino may remain
massless or gain a non--zero mass.

The main disadvantage of the scenario considered above is related
with the degeneracy between bosons and fermions in the observable
sector, which is preserved in the limit $\xi_{\beta},
\eta_{\alpha\beta}\to 0$ despite the breakdown of local
supersymmetry. In the general case, when both
$\xi_{\alpha}$ and $\zeta_{\alpha}$ have non--zero values,
the situation changes dramatically. Since, by construction,
the dilatation invariance is only broken in the part of the
K$\Ddot{a}$hler potential (\ref{192}) containing observable
superfields, it does not affect the scalar potential of the
hidden sector which is still described by Eq.~(\ref{193}).
As a result our MPP scenario is realized without any extra
fine--tuning.

Nevertheless the shape of the effective scalar potential of the
observable sector, in the vacua where the super-Higgs effect takes
place, alters significantly. The structure of the soft SUSY breaking 
terms in the considered model, which is discussed in the Appendix,
allows us to write the effective potential of the observable
sector (\ref{195}) in a compact form\footnote{This form
of the scalar potential can be established in a straightforward
way in the limit when all $\zeta_{\alpha}$ go to zero
($x_{\alpha}\to\infty$). Then the K$\Ddot{a}$hler metric of
the observable superfields $K_{\bar{\alpha}\beta}$ is diagonal
and does not depend on the hidden sector superfields, which makes
the computation of the SUGRA scalar potential relatively simple.}:
\be
V_{eff}(\varphi_{\alpha}, \varphi^{*}_{\alpha})\simeq
\sum_{\alpha}\biggl|\ds\frac{\partial
W_{eff}(y_{\beta})}{\partial y_{\alpha}}
+m_{\alpha}y^{*}_{\alpha}\biggr|^2+\ds\frac{1}{2}\sum_{a}(D^{a})^2\,.
\label{199}
\ee
Although global supersymmetry is softly broken, the effective
potential of the scalar fields (\ref{199}) is still positive
definite and vanishes near its global minima. It follows
that the spontaneous breakdown of electroweak symmetry can not
be naturally arranged in our model, because normally it results
in negative vacuum energy density, i.e.~the minimum of the
scalar potential with broken $SU(2)_W\times U(1)_Y$ symmetry
ought to be deeper than the vacuum where gauge invariance is preserved
and the doublet Higgs fields vanish $(<H_1>=<H_2>=0)$.
Moreover in the simplest MPP inspired SUGRA model
discussed above, the mechanism for the stabilization of the
vacuum expectation value of the hidden sector field $T$ remains
unclear. As a result the gravitino mass (see Eq.~(\ref{194}))
and the supersymmetry breaking scale are not fixed in the physical vacuum.

However all these problems can not be addressed in the framework
of the simplest MPP inspired SUGRA model. In order to get a
self--consistent solution, one has to include all perturbative
and non--perturbative corrections to the considered SUGRA Lagrangian,
which should depend on the structure of the underlying theory.
If we take into account the evolution of the soft scalar masses,
then their renormalization group flow might provide a radiative
mechanism for electroweak symmetry breaking \cite{36}. We hope that
an underlying renormalizable or even finite theory can be found,
which sheds light on the origin of the terms that spoil the
global symmetry protecting the cosmological constant in our SUGRA
model. It should also ensure the stabilization of the vacuum
expectation values of the hidden sector fields and the supersymmetry
breaking scale.

\section{Cosmological constant in MPP inspired SUGRA models}
\label{cosmological}

We now assume that a phenomenologically viable MPP inspired SUGRA
model of the type just discussed can be constructed. That is to
say, we assume the existence of a phase with electroweak gauge
symmetry breaking induced by soft SUSY breaking terms degenerate
with a second phase, in which the low--energy limit of the
considered theory is described by a pure supersymmetric model in
flat Minkowski space. Non-perturbative effects in the observable
sector may lead to supersymmetry breakdown in the second vacuum
state (for recent reviews see \cite{26}--\cite{27}). Then in
compliance with our MPP philosophy, we require the degeneracy of
the vacua after all non-perturbative effects are included.

The non-perturbative effects in simple SUSY models, like the
minimal supersymmetric standard model (MSSM), are extremely weak.
Our strategy is to estimate these effects in the second vacuum and
thereby estimate the energy density in the second (almost
supersymmetric) phase. This value of the cosmological constant
can then be interpreted as the physical value in our phase, by
virtue of MPP.

If supersymmetry breaking takes place in the second vacuum, it is
caused by the strong interactions. Indeed, even in the pure MSSM,
the beta function of the strong gauge coupling constant $\alpha_3$
exhibits asymptotically free behaviour ($b_3=-3$)\footnote{ The
gauge couplings obey the renormalization group equations
$\ds\frac{d\log{\alpha_i(Q)}}{d\log{Q^2}}=\frac{b_i\alpha_i(Q)}{4\pi}$,
where $\alpha_i(Q)=g_i^2(Q)/(4\pi)$.}. As a consequence
$\alpha_3(Q)$ increases in the infrared region and one can expect
that the role of non--perturbative effects is enhanced. Since in
the minimal SUGRA model the kinetic functions essentially do not
depend on the hidden superfields ($f_a(z_m) \simeq const$), the
values of the gauge couplings at the high energy scale and their
running down to the scale $M_{S}\simeq m_{3/2}$ are the same in
both vacua. Below the scale $M_S$ all superparticles in the
physical vacuum decouple and the corresponding beta functions
change ($\tilde{b}_3=-7$). Using the value of
$\alpha^{(1)}_3(M_Z)\approx 0.118\pm 0.003$ and the matching
condition $\alpha^{(2)}_3(M_S)=\alpha^{(1)}_3(M_S)$, one finds the
strong coupling in the second vacuum 
\be
\ds\frac{1}{\alpha^{(2)}_3(M_S)}=\ds\frac{1}{\alpha^{(1)}_3(M_Z)}-
\frac{\tilde{b}_3}{4\pi}\ln\frac{M^2_{S}}{M_Z^2}\, . 
\label{22}
\ee 
In Eq.(\ref{22}) $\alpha^{(1)}_3$ and $\alpha^{(2)}_3$ are the
values of the strong gauge couplings in the physical and second
minima of the SUGRA scalar potential.

At the scale $\Lambda_{SQCD}$, where the supersymmetric QCD
interactions become strong in the second vacuum
\be
\Lambda_{SQCD}=M_{S}\exp\left[{\frac{2\pi}{b_3\alpha_3^{(2)}(M_{S})}}\right]\,
\label{23}
\ee
the supersymmetry may be broken dynamically due to
non--perturbative effects. If instantons generate a repulsive
superpotential \cite{26}, \cite{28}--\cite{281} which lifts and
stabilizes the vacuum valleys in the scalar potential, then a
generalized O'Raifeartaigh mechanism gives rise to a non--zero
positive value for the cosmological constant
\be
\Lambda \simeq \Lambda_{SQCD}^4\, .
\label{24}
\ee

In Fig.~1 the dependence of $\Lambda_{SQCD}$ on the SUSY breaking
scale $M_S$ is examined. Because $\tilde{b}_3 < b_3$ the QCD gauge
coupling below $M_S$ is larger in the physical minimum than in the
second one. Therefore the value of $\Lambda_{SQCD}$ is much lower
than the QCD scale in the Standard Model and diminishes with
increasing $M_S$. When the supersymmetry breaking scale in our
vacuum is of the order of 1 TeV, we obtain
$\Lambda_{SQCD}=10^{-26}M_{Pl} \simeq 100$ eV. This results in an
enormous suppression of the total vacuum energy density
($\Lambda\simeq 10^{-104} M_{Pl}^4$) compared to say an
electroweak scale contribution in our vacuum $v^4 \simeq 10^{-62}
M_{Pl}$.  From the rough estimate of the energy density
(\ref{24}), it can be easily seen that the measured value of the
cosmological constant is reproduced when
$\Lambda_{SQCD}=10^{-31}M_{Pl} \simeq 10^{-3}$ eV. The appropriate
values of $\Lambda_{SQCD}$ can therefore only be obtained for
$M_S=10^3-10^4\,\mbox{TeV}$. However the consequent large
splitting within SUSY multiplets would spoil gauge coupling
unification in the MSSM and reintroduce the hierarchy problem,
which would make the stabilization of the electroweak scale rather
problematic.

A model consistent with electroweak symmetry breaking and
cosmological observations can be constructed, if the MSSM particle
content is supplemented by an additional pair of $5+\bar{5}$
multiplets. These new bosons and fermions would not affect gauge
coupling unification, because they form complete representations
of $SU(5)$ (see for example \cite{29}). In the physical vacuum
these extra particles would gain masses around the supersymmetry
breaking scale. The corresponding mass terms in the superpotential
are generated after the spontaneous breaking of local
supersymmetry, due to the presence of the bilinear terms
$\left[\eta (5\cdot \overline{5})+h.c.\right]$ in the
K$\Ddot{a}$hler potential of the observable sector \cite{30}. Near
the second minimum of the SUGRA scalar potential the new particles
would be massless, since $m_{3/2}=0$. Therefore they give a
considerable contribution to the $\beta$ functions ($b_3=-2$),
reducing $\Lambda_{SQCD}$ further. In this case the observed value
of the cosmological constant can be reproduced even for $M_S\simeq
1\,\mbox{TeV}$ (see Fig.~1).

Unfortunately achieving dynamical SUSY breaking at the scale
$\Lambda_{SQCD}$ is actually not at all easy. The situation is different
depending on whether the number of flavours $N_f$ is larger or smaller than
the number of colours $N_c$. In the MSSM and its simplest extensions, where
$N_c=3$ and $N_f=6$, the generated superpotential has a polynomial
form \cite{27}, \cite{31}. The absolute minimum of the SUSY
scalar potential is then reached when all the superfields,
including their F- and D-terms, acquire zero vacuum expectation
values preserving supersymmetry in the second vacuum. This result
throws some doubt on our estimations of the value of the cosmological constant,
which is based on Eq.~(\ref{24}).

But the above disappointing facts concerning dynamical SUSY
breaking were revealed in the framework of pure supersymmetric
QCD, where all Yukawa couplings were supposed to be small or even
absent. At the same time the t--quark Yukawa coupling in the MSSM
is of the same order of magnitude as the strong gauge coupling at
the electroweak scale. Therefore it might change the results of
the SUSY breaking studies drastically leading, for example, to the
formation of a quark condensate that breaks supersymmetry.

\section{Implementation of the MPP in the models with extended gauge symmetry}
\label{extended}

The breakdown of supersymmetry in the observable sector can be
more easily achieved in models with an extended strong interaction
gauge sector. Here we restrict our consideration to the class of
models based on $SU(N)$ gauge symmetry groups. Since the extension
of the gauge sector of the SM is already a very strong assumption,
we prefer to limit the particle content of the model as much as
possible. In particular it is worth combining the spontaneous
breakdowns of the enlarged gauge symmetry and local supersymmetry,
as takes place in our toy $SU(5)$ model (\ref{12}), rather than
introducing two separate sectors for this purpose. It seems that
the simplest gauge extension of the MSSM, for which the dynamical
supersymmetry breaking occurs at low energies independently of the
values of the Yukawa couplings, should include at least three
$SU(3)$ gauge groups. If the quarks of each generation are coupled
to the gauge bosons of just their own distinct $SU(3)$, then the
criterion $N_c>N_f$ is satisfied. We here consider a model with an
$\biggl[SU(3)\biggr]^3$ gauge symmetry, as in the family
replicated gauge group or anti-grand unification model \cite{10},
\cite{itep}. As in this model, we take the corresponding three
$\biggl[SU(3)\biggr]^3$ gauge coupling constants to be equal and
we denote their value at the scale $Q$ by $g_{33}(Q)$.

In the physical vacuum the $\biggl[SU(3)\biggr]^3$ gauge symmetry
must be broken down to $SU(3)_C$, which is associated with the SM
strong interactions. This can be simply arranged if the considered
theory includes multiplets in the bi--fundamental representation,
which transform as a triplet with respect to one $SU(3)$ and as an
anti--triplet under another SU(3) symmetry. In the models based on
$SU(3)_a\times SU(3)_b\times SU(3)_c$ there can be six
bi--fundamental representations: $\Phi_{a\bar{b}}$,
$\Phi_{a\bar{c}}$, $\Phi_{b\bar{a}}$, $\Phi_{b\bar{c}}$,
$\Phi_{c\bar{a}}$, $\Phi_{c\bar{b}}$ where the indices $a$, $b$
and $c$ correspond to the three different $SU(3)$ gauge groups and
the corresponding quark generations. If the superfields
$\Phi_{i\bar{j}}$ acquire vacuum expectation values
\be
\Phi_{i\bar{j}}=\Phi_0\left( \ba{ccc}
~1~ & ~0~ & ~0~\\[2mm]
~0~ & ~1~ & ~0~\\[2mm]
~0~ & ~0~ & ~1~ \ea \right)\,,\qquad i,j=a,b,c
\label{25}
\ee
then below the energy scale $\Phi_0$ the $\biggl[SU(3)\biggr]^3$ gauge
group reduces to the diagonal subgroup corresponding to the usual
QCD $SU(3)_C$ symmetry. It follows that the QCD gauge coupling
constant $g_3(Q)$ is then related to the $\biggl[SU(3)\biggr]^3$
gauge coupling constant at the scale $\Phi_0$:
\be
g_{33}^{(1)}(\Phi_0+\varepsilon) = \sqrt{3}g_3^{(1)}(\Phi_0-\varepsilon).
\label{29}
\ee

The desired pattern of $\biggl[SU(3)\biggr]^3$ gauge symmetry
breaking can be obtained in the no--scale SUGRA model with
superpotential
\be
\ba{c}
W=\mu_X\biggl[\mbox{Tr}\left(\Phi_{a\bar{b}}\Phi_{b\bar{a}}\right)
+\mbox{Tr}\left(\Phi_{a\bar{c}}\Phi_{c\bar{a}}\right)
+\mbox{Tr}\left(\Phi_{b\bar{c}}\Phi_{c\bar{b}}\right)\biggr]
+\qquad\qquad\qquad\qquad\qquad\\[2mm]
\qquad\qquad\qquad\qquad\qquad\qquad\qquad
+k\biggl[\mbox{Tr}\left(\Phi_{a\bar{b}}\Phi_{b\bar{c}}\Phi_{c\bar{a}}\right)
+\mbox{Tr}\left(\Phi_{b\bar{a}}\Phi_{a\bar{c}}\Phi_{c\bar{b}}\right)\biggr]
+\hat{W}(\varphi_{\sigma})\,.
\ea
\label{26}
\ee
and K$\ddot{a}$hler potential
\be
K(\phi_{M},\phi_{M}^{*})=\ds-3\ln\biggl[T+\overline{T}
-\sum_{i,j}|\Phi_{i\bar{j}}|^2\biggr]
+\hat{K}(\varphi_{\sigma},\varphi_{\sigma}^{*})
\label{261}
\ee
where $\hat{W}(\varphi_{\sigma})$ and
$\hat{K}(\varphi_{\sigma},\varphi_{\sigma}^{*})$ depend on the
Higgs, quark and lepton superfields $\varphi_{\sigma}$. The model
possesses two degenerate minima, where
$\Phi_0=-\ds\frac{\mu_X}{k}$ and $\Phi_0=0$ respectively. In the
first vacuum $\left(\Phi_0=-\ds\frac{\mu_X}{k}\right)$ local
supersymmetry and $\biggl[SU(3)\biggr]^3$ gauge symmetry are
broken. The breaking of global supersymmetry can be induced at low
energies as well, if the part of the K$\ddot{a}$hler potential 
$\hat{K}(\varphi_{\sigma},\varphi_{\sigma}^{*})$ is not invariant
under the dilatation transformations. For example, in the simple
case considered where $\hat{K}$ does not depend on $T$ and
$\Phi_{i\bar{j}}$, i.e.
\be
\hat{K}(\varphi_{\sigma},\varphi_{\sigma}^{*})=\sum_{\alpha,
\beta}\biggl(\ds\frac{\eta_{\alpha\beta}}{2}\,\varphi_{\alpha}\,
\varphi_{\beta}+h.c.\biggr)+ \sum_{\sigma}
\xi_{\sigma}|\varphi_{\sigma}|^2\,,
\label{27}
\ee
the scalar components of the observable superfields
$\varphi_{\sigma}$ gain a universal mass which coincides with the
gravitino mass $m_{3/2}=\ds\frac{3\mu_X^3}{k^2<(T+\overline{T}
-18(\mu_X/k)^2)^{3/2}>}$ (see Eq.~(\ref{198}) where 
$x_{\alpha}\to\infty$). For simplicity, we take $k$ of order unity 
in the following.

In the second vacuum ($\Phi_0=0$) supersymmetry and gauge
symmetries are left unbroken. The gauge couplings of each
$SU(3)_i$ grow with decreasing energy scale developing a Landau
pole much below $\mu_X$. At low energies, where the $SU(3)_i$
gauge interactions become very strong ($E\simeq \Lambda_{SQCD}$),
non--perturbative effects induce a sizable instanton contribution
$W_{inst}$ to the effective superpotential (see \cite{28}) that
takes the form
\be
\ba{rcl}
W&=&W_{inst}+h_t(\hat{H}_{u}\hat{Q})\hat{t}^c
+h_b(\hat{H}_{d}\hat{Q})\hat{b}^c
+h_{\tau}(\hat{H}_{d}\hat{L})\hat{\tau}^c\,,\\[2mm]
W_{inst}&\simeq&\ds\frac{\Lambda^7_{SQCD}}
{(Q_{\alpha}\,t^c)\epsilon_{\alpha\beta}(Q_{\beta}\,b^c)}\,.
\ea
\label{28}
\ee
For simplicity we only keep superfields belonging
to the third generation in the superpotential (\ref{28}), together
with the Higgs doublets $H_u$ and $H_d$. In Eq.~(\ref{28})
$\alpha$ and $\beta$ are $SU(2)$ indices labelling the components
of the $SU(2)$ doublet $Q_{\alpha}$, whereas
$\epsilon_{\alpha\beta}$ is the completely antisymmetric tensor.
The non--perturbative superpotential $W_{inst}$ gives rise to
supersymmetry breaking. Indeed, in a vacuum where supersymmetry is
preserved, all the auxiliary fields $F_i$ have to be zero. The
vanishing of $F_{H_u}$ implies that the vacuum expectation value
of either $<Q>$ or $<t^c>$ is zero. At the same time with
the superpotential (\ref{28}), $W_{inst}$ as well as $F_{Q}$ and
$F_{t^c}$ are singular when $<Q>=0$ or $<t^c>=0$. Therefore it is
not consistent to assume that supersymmetry is preserved in the
vacuum, but non-perturbative instanton effects must break the
supersymmetry and give rise to a non--zero vacuum energy density
$\Lambda\simeq \Lambda_{SQCD}^4$.

So far the gauge kinetic function in the considered model has not
been specified. In contrast with the simplest MPP inspired models,
a constant gauge kinetic function in this particular gauge
extension of the SM does not allow us to reproduce the observed
value of the cosmological constant. In realistic scenarios the
supersymmetry breaking scale in the physical vacuum has to be
above a few hundred GeV, restricting the permitted range of
$\mu_X$ from below. Assuming that $T$ gets a vacuum expectation
value around unity (i.e. $T \sim M_{Pl}$), the scale of
$\biggl[SU(3)\biggr]^3$ symmetry breaking ought to be higher than
$10^{13}\,\mbox{GeV}$ but should not exceed $M_{Pl}$. In order to
get a phenomenologically acceptable value for the vacuum energy
density in the second minimum, which according to MPP coincides
with the cosmological constant in our vacuum, we require
$\Lambda_{SQCD} \simeq 10^{-3}$ eV. Hence the $SU(3)$ gauge
couplings $g_{33}^{(2)}$ at the scale $\mu_X$ are required to be
in the vicinity of $0.4$ in the second vacuum (see Fig.~2).
However, the value of the $SU(3)$ gauge couplings in the physical
vacuum $g_{33}^{(1)}$ just above the scale $\mu_X$ is considerably
larger than $g_{33}^{(2)}(\mu_X)$, as one can see from Fig.~3\,.

Thus, in order to obtain an appropriate value of $\Lambda_{SQCD}$,
the $SU(3)_i$ gauge couplings in the second vacuum have to be two
or three times smaller than in the physical one. This can be
achieved if the gauge kinetic function depends quite strongly on
the vacuum expectation values of the bi--fundamental multiplets
$\Phi_{i\bar{j}}$. The simplest gauge kinetic function for the
gauge group $SU(3)_a$, which is invariant under gauge symmetry
transformations, imaginary translations and dilatations, can be
written as 
\be 
f_{a}(\phi_M)=f_{a}^0+\sum_{i,j}f^{a}_{i\bar{j}}
\frac{|\Phi_{i\bar{j}}|^2}{(T+\bar{T})}\,. 
\label{30} 
\ee 
When we take  $f_{a}^0\simeq 6.28$, i.e.~ $(g_{33}^{(2)}(M_{Pl}))^2
=1/f_a^0 \simeq 0.16$, the gauge couplings of
$\biggl[SU(3)\biggr]^3$ blow up near the scale
$\Lambda_{SQCD}\simeq 10^{-3}\,\mbox{eV}$, inducing a suitable
value of the vacuum energy density. In the physical vacuum the
gauge couplings $g_{33}^{(1)}(M_{Pl})$ differ from
$g_{33}^{(2)}(M_{Pl})$, because the bi--fundamental multiplets
acquire non--zero vacuum expectation values. If the second term in
Eq.~(\ref{30}) takes the value $(-4.9)$ in the physical vacuum,
the measured value of $\alpha^{(1)}_3(M_Z)$ is reproduced using
Eq.~(\ref{29}). This value can be obtained with all the parameters
$f_{a}^0$ and $f^{a}_{i\bar{j}}$ of the same order of magnitude,
provided that $\Phi_0\simeq M_{Pl}$.

In the case when $\Phi_0\simeq M_{Pl}$ the gauge symmetry, global
and local supersymmetry are all broken just below the Planck scale
in the physical vacuum. As can be seen from Fig.~3, the
$\biggl[SU(3)\biggr]^3$ gauge couplings then take the value
$g_{33}^{(1)}(M_{Pl})\simeq 0.85$. This is consistent with the
critical value of the gauge coupling constant obtained from
lattice calculations \cite{zuber}, for which three phases of the
regularised $SU(3)$ gauge theory coexist, i.e. for which the
corresponding vacuum states have the same energy density in
agreement with our MPP philosophy. Similar results were obtained
for the $\biggl[SU(2)\biggr]^3$ and $\biggl[U(1)\biggr]^3$ gauge
couplings in the family replicated gauge group model \cite{10},
\cite{itep}, using the measured values of $\alpha_2(M_Z)$ and
$\alpha_1(M_Z)$ as inputs. We note that a phenomenologically
successful structure for the quark and lepton mass matrices can be
naturally generated from the chiral gauge charges in the family
replicated gauge group model \cite{32}.

\section{Summary and concluding remarks}
\label{conclusion}

In supergravity the cosmological constant problem can be
alleviated by imposing an extra global symmetry. In particular the
invariance under imaginary translations and dilatations, which are
subgroups of $SU(N,1)$, leads to the vanishing of the vacuum
energy density in the no-scale SUGRA models. At the same time
these symmetries, which naturally arise in theories with extended
supersymmetry ($N\ge 5$), preserve local supersymmetry which must
however be broken in any phenomenologically acceptable theory. We
have argued that the breakdown of these global symmetries
protecting the cosmological constant does not necessarily result
in a non--zero vacuum energy density. In particular, violation of
dilatation invariance in the superpotential of no--scale models
may give rise to the spontaneous breakdown of local supersymmetry,
and still preserve a zero value for the energy density in the
vacua of these models.

All global minima of the SUGRA scalar potential (\ref{2}) in the
no--scale models, where the invariance with respect to dilatations
is spoiled in the superpotential, are degenerate. Normally the set
of global minima in the considered class of models includes vacua
with broken and unbroken local supersymmetry. In the vacua where
local supersymmetry remains intact, the gravitino mass goes to
zero and the conditions (\ref{15}) are fulfilled automatically.
According to our MPP scenario the SUGRA scalar potential must
possess at least two degenerate vacua in which $m_{3/2}=0$ and
$m_{3/2}\ne 0$ respectively. In one of them, where $m_{3/2}$ has a
non--zero value, local supersymmetry is broken in the hidden
sector at the high energy scale $(\sim 10^{10}-10^{12}\,
\mbox{GeV})$, inducing a set of soft SUSY breaking terms for the
observable fields. In the other vacuum ($m_{3/2}=0$) the low
energy limit of the considered theory is described by a pure
supersymmetric model in flat Minkowski space. The energy density
and all auxiliary fields $F^M$ of the hidden sector vanish in this
second vacuum preserving supersymmetry.

Although the breakdown of dilatation invariance in the
superpotential of no--scale SUGRA models ensures the degeneracy of
vacua, where $m_{3/2}=0$ and $m_{3/2}\ne 0$ respectively, the
particle spectrum remains supersymmetric at low energies in all
vacua. Thereby none of these vacua can be the physical one.
Nevertheless a minimal SUGRA model has been constructed, where our
MPP scenario is realized without any extra fine--tuning. It is
based on broken $SU(N,1)$ symmetry. The hidden sector of the
minimal MPP inspired SUGRA model contains two superfields $T$ and
$z$, which transform differently under imaginary translations and
dilatations. We allowed the breakdown of dilatation invariance in
the superpotential of the hidden sector and in the part of the
K$\ddot{a}$hler potential which contains the observable
superfields. The $SU(N,1)$ structure of the K$\ddot{a}$hler
potential of the hidden sector guarantees the vanishing of the
cosmological constant in all the global minima of the scalar
potential in the model. Owing to the breakdown of dilatation
invariance in the hidden sector superpotential, a set of
degenerate vacua with broken and unbroken local supersymmetry
emerges. Meantime we maintain dilatation invariance in the
observable sector superpotential, preventing the appearance of
bilinear and high order terms involving observable superfields in
the rest of the superpotential and thereby eliminating the
$\mu$--problem. Finally, due to a suitable breakdown of dilatation
invariance in the K$\ddot{a}$hler potential of the observable
sector, effective $\mu$--terms and a set of soft SUSY breaking
terms are generated in the vacua where local supersymmetry is
spontaneously broken. 

In spite of the vanishing of the vacuum energy density in all
global minima of the tree level scalar potential of the MPP
inspired SUGRA models, the value of the cosmological constant may
differ from zero. This occurs if non--perturbative effects in the
observable sector give rise to the breakdown of supersymmetry in
the second vacuum (phase). Our MPP philosophy then requires that
the phase in which local supersymmetry is broken in the hidden
sector has the same energy density as a phase where supersymmetry
breakdown takes place in the observable sector. If the gauge
couplings at high energies are identical in both vacua, the value
of the energy density in the second vacuum can be estimated
relatively easily. It is positive definite and determined by the
scale where the $SU(3)_C$ gauge interactions become strong. The
numerical analysis carried out in the framework of the pure MSSM
has revealed that the corresponding scale is naturally low
($\Lambda_{SQCD}\lesssim 10^{-25}\,M_{Pl}$) for a reasonable
choice of the supersymmetry breaking scale, $M_S\gtrsim
1\,\mbox{TeV}$, in the first (physical) vacuum. Moreover the
introduction of an extra pair of $5+\bar{5}$ multiplets reduces
this scale down further, so that the energy density of the second
phase approaches the observed value of the cosmological constant
even when $M_S\simeq 1\,\mbox{TeV}$. The crucial idea is then to
use MPP to transfer the energy density or cosmological constant
from this second vacuum into all other vacua, especially into the
physical one in which we live. In such a way we have suggested an
explanation of why the observed value of the cosmological constant
is positive and takes on the tiny value it has. The MPP scenario
with additional $5+\bar{5}$ multiplets of matter and supersymmetry
breaking scale in the $\mbox{TeV}$ range can be tested at the LHC
or ILC.

The trouble with the MPP prediction for the value of the
cosmological constant is that it is not clear if the required
dynamical supersymmetry breaking actually takes place in the
framework of the simplest SUSY extensions of the SM, which
describe the observable sector of SUGRA models at low energies. On
the other hand, the dynamical breakdown of supersymmetry can be
attained in SUSY models with an extended gauge sector for the
strong interactions similar to that in the family replicated gauge
group model \cite{10}, \cite{itep}, \cite{32}. But, in order to
obtain the appropriate value of the cosmological constant in this
case, the gauge couplings in the first and second vacua should
differ considerably. Therefore one has to admit a dependence of
the gauge kinetic function on the chiral superfields, which are
responsible for the breaking of the enlarged gauge symmetry down
to $SU(3)_C\times SU(2)_W\times U(1)_Y$. Then, if local
supersymmetry and the extended gauge symmetry are broken near the
Planck scale, the gauge couplings in the second vacuum can be
smaller than in the physical one by a factor of 2, which allows us
to reproduce the observed value of the cosmological constant. In
the first vacuum where we live the SM is valid up to the Planck
scale. It has recently been pointed out that the enormous
hierarchy between the electroweak and Planck scales might also be
explained by MPP \cite{itep}, \cite{33} in the SM.

Although MPP provides an attractive explanation for the smallness
and sign of the cosmological constant in $(N=1)$ supergravity, we
have not been able to present a fully self--consistent model. The
no--scale models discussed above possess one defect. Namely, the
mechanism for the stabilization of the vacuum expectation value of
the hidden sector field $T$ and the SUSY breaking scale remains
unclear.

\vspace{-2mm}

\section*{Acknowledgements}

\vspace{-2mm} The authors are grateful to O.~Kancheli, S.~King
and D.~Sutherland for fruitful discussions. RN would
like to acknowledge support from the PPARC grant PPA/G/S/2003/00096.
RN was also partly supported by a Grant of the President of Russia
for young scientists (MK--3702.2004.2). The work of CF was
supported by PPARC and the Niels Bohr Institute Fund. CF would
like to acknowledge the hospitality of the Niels Bohr Institute,
while part of this work was done.

\newpage

\section*{Appendix}

Here we discuss the structure of the soft SUSY breaking terms which
appear in the physical vacuum in a low energy effective Lagrangian 
of the MPP inspired SUGRA model with superpotential (\ref{191})
and K$\Ddot{a}$hler potential given by Eq.~(\ref{192}).
In order to compute the effective scalar potential, 
one has to substitute vacuum expectation values for $T$ and $z$ 
as well as for their auxiliary fields (\ref{3}), taking into 
account that only $F^{T}$ acquires a non--zero vacuum expectation 
value. Then one expands the full SUGRA scalar potential (\ref{1}) 
in powers of observable fields, taking the flat limit \cite{21} 
where $M_{Pl}\to\infty\,$ but $m_{3/2}$ is kept fixed. In the 
considered limit hidden sector superfields are decoupled from 
the low--energy theory. The only signal they produce is a set 
of terms that break the global supersymmetry of the low--energy 
effective Lagrangian of the observable sector in a soft 
way \cite{7}, \cite{8}, i.e. without inducing quadratic divergences. 
All non--renormalizable terms vanish in the flat limit since they 
are suppressed by inverse powers of $M_{Pl}$. Thus one is left 
with a global SUSY scalar potential $V_{SUSY}$ plus a set of soft 
SUSY breaking terms $V_{soft}$, i.e.
\be
\ba{c}
V_{eff}(y_{\alpha}, y^{*}_{\alpha})= V_{SUSY}+V_{soft}\,,\\[3mm]
V_{SUSY}=\sum_{\alpha}\biggl|\ds\frac{\partial W_{eff}(y_{\beta})}
{\partial y_{\alpha}}\biggr|^2+\ds\frac{1}{2}\sum_{a}(D^{a})^2\,,\\[3mm]
V_{soft}=\sum_{\alpha}m^2_{\alpha}|y_{\alpha}|^2
+\biggl[\sum_{\alpha,\,\beta}\ds\frac{1}{2}B_{\alpha\beta}
\mu_{\alpha\beta}y_{\alpha}y_{\beta}+
\sum_{\alpha,\,\beta,\,\gamma}\ds\frac{1}{6}A_{\alpha\beta\gamma}
h_{\alpha\beta\gamma}y_{\alpha}y_{\beta}y_{\gamma}+h.c.\biggr]\,.
\ea
\label{195}
\ee
In Eq.~(\ref{195}) $y_{\alpha}$ are canonically normalized scalar
fields
\be
\ba{c}
y_{\alpha}=\tilde{C}_{\alpha}\varphi_{\alpha}\,,\qquad
\tilde{C}_{\alpha}=\xi_{\alpha}\biggl(1+\ds\frac{1}
{x_{\alpha}}\biggr)\,,\qquad x_{\alpha}
=\frac{\xi_{\alpha}<(T+\overline{T}-|z|^2)>}{3\zeta_{\alpha}}\,.
\ea
\label{196}
\ee
When $M_{Pl}\to\infty\,$ the effective superpotential, which
describes the interactions of observable superfields
at low energies, only contains bilinear and trilinear terms
\be
\ba{c}
W_{eff}=\sum_{\alpha,\,\beta}\ds\frac{\mu_{\alpha\beta}}{2}\,
\varphi_{\alpha}\,\varphi_{\beta}+\sum_{\alpha,\,\beta,\,\gamma}
\ds\frac{h_{\alpha\beta\gamma}}{6}\,\varphi_{\alpha}\,
\varphi_{\beta}\,\varphi_{\gamma}\,,\\[3mm]
\mu_{\alpha\beta}=m_{3/2}\eta_{\alpha\beta}(\tilde{C}_{\alpha}
\tilde{C}_{\beta})^{-1}\,,\qquad\qquad
h_{\alpha\beta\gamma}=\ds\frac{Y_{\alpha\beta\gamma}
(\tilde{C}_{\alpha}\tilde{C}_{\beta}\tilde{C}_{\gamma})^{-1}}
{<(T+\overline{T}-|z|^2)^{3/2}>}\,.
\ea
\label{197}
\ee

The complete set of soft SUSY breaking terms involves:
gaugino masses $M_a$, masses of scalar components of
observable superfields $m_{\alpha}$,
trilinear $A_{\alpha\beta\gamma}$ and bilinear
$B_{\alpha\beta}$ scalar couplings associated with
Yukawa couplings and $\mu$--terms in the
superpotential \cite{23}. Three types of soft SUSY
breaking parameters $m^2_{\alpha}$, $A_{\alpha\beta\gamma}$
and $B_{\alpha\beta}$ appear in the scalar potential
(\ref{195}). In the vacua, where local SUSY is broken
and the gravitino gains a non-zero mass $m_{3/2}$, these
parameters are given by\footnote{In the most general case
a complete set of expressions for the soft SUSY breaking
parameters can be found in \cite{24}--\cite{25}.}
\be
\ba{rcl}
m_{\alpha}&=&\ds m_{3/2}\frac{x_{\alpha}}{(1+x_{\alpha})}\,,\\[3mm]
B_{\alpha\beta}&=&m_{\alpha}+m_{\beta}\,,\\[3mm]
A_{\alpha\beta\gamma}&=&m_{\alpha}+m_{\beta}+m_{\gamma}\,.
\ea
\label{198}
\ee
The structure of the soft SUSY breaking terms given above permits 
to rewrite the effective scalar potential (\ref{195})
in a more compact form (\ref{199}).
It is worth emphasizing that the expressions for the soft SUSY
breaking parameters obtained above would not change if the hidden 
sector of our model had many superfields $z_i$. The soft scalar 
masses $m_{\alpha}$ in the low energy effective Lagrangian maintain 
the splitting between bosons and fermions within one supermultiplet. 
According to Eq.~(\ref{198}), the masses of the superpartners of the
ordinary quarks and leptons are set by the parameter
$\xi_{\alpha}$ and the vacuum expectation value of the superpotential
of the hidden sector (or $\varkappa$), which spoil
the dilatation invariance. In other words the qualitative pattern
of the sparticle spectrum in the considered SUGRA model depends
on the extent to which the symmetry protecting the cosmological
constant is broken. Assuming that $\xi_{\alpha}$, $\zeta_{\alpha}$, $\mu_0$ 
and $<T>$ are all of order unity, the phenomenologically acceptable 
value of the supersymmetry breaking scale $M_S\sim 1\,\mbox{TeV}$ can
only be obtained for extremely small values of
$\varkappa\simeq 10^{-15}$. 

Explicit expressions for the gaugino masses are not included in
Eq.~(\ref{198}) because their values are determined by the
gauge kinetic functions $f_a(T,\,z)$ that has not been specified
yet. A canonical choice for the kinetic function in minimal
supergravity $f_a(T,\, z)=const$ corresponds to $M_a=0$.
In order to avoid a conflict with chargino and gluino searches
at present and former colliders, we need
gaugino masses in the few hundred $\mbox{GeV}$ range. Therefore
we assume a mild dependence of $f_a(T,\, z)$ on the hidden sector
fields, which is strong enough to induce appreciable gaugino
masses but weak enough to ensure that the gauge couplings in
the physical and supersymmetric vacua do not differ significantly.

\newpage
\noindent
{\Large \bf Figure captions}

\noindent {\bf Fig. 1.}\, The value of
$\log\left[\Lambda_{SQCD}/M_{Pl}\right]$ versus $\log M_S$. The
thin and thick solid lines correspond to the pure MSSM and the
MSSM with an additional pair of $5+\bar{5}$ multiplets
respectively. The dashed and dash--dotted lines represent the
uncertainty in $\alpha_3(M_Z)$. The upper dashed and dash-dotted
lines are obtained for $\alpha_3(M_Z)=0.124$, while the lower ones
correspond to $\alpha_3(M_Z)=0.112$. The horizontal line
represents the observed value of $\Lambda^{1/4}$. The SUSY
breaking scale $M_S$ is measured in GeV.

\vspace{5mm} \noindent {\bf Fig. 2.}\, The value of the vacuum
energy density as a function of the overall
$\biggl[SU(3)\biggr]^3$ gauge coupling at the scale $\mu_X$ in the
second vacuum. The dash-dotted and thick curves represent the
dependence of the energy density on $g_{33}^{(2)}(\mu_X)$ for
$\mu_X=M_{Pl}$ and $\mu_X=10^{13}\,\mbox{GeV}$ respectively. The
horizontal solid line corresponds to the observed value of the
cosmological constant $\Lambda$.

\vspace{5mm} \noindent {\bf Fig. 3.}\, The dependence of the
overall $\biggl[SU(3)\biggr]^3$ gauge coupling $g_{33}(\mu_X)$ on
the scale $\mu_X$. The upper solid curve represents
$g_{33}^{(1)}(\mu_X)$ and is obtained by the extrapolation of
$\alpha_3(M_Z)$ up to the scale $\mu_X$ in the physical vacuum.
The lower thick line represents the values of
$g_{33}^{(2)}(\mu_X)$ that allow us to fit the vacuum energy
density in the second vacuum to its phenomenologically acceptable
value $\Lambda \simeq 10^{-123}M_{Pl}^4$. The scale of the
$\biggl[SU(3)\biggr]^3$ symmetry breaking $\mu_X$ is given in GeV.

\newpage
\noindent
\hspace{-1cm}{\Large $\log[\Lambda_{SQCD}/M_{Pl}]$}\\
\begin{center}
{\hspace*{-20mm}\includegraphics[height=90mm,keepaspectratio=true]{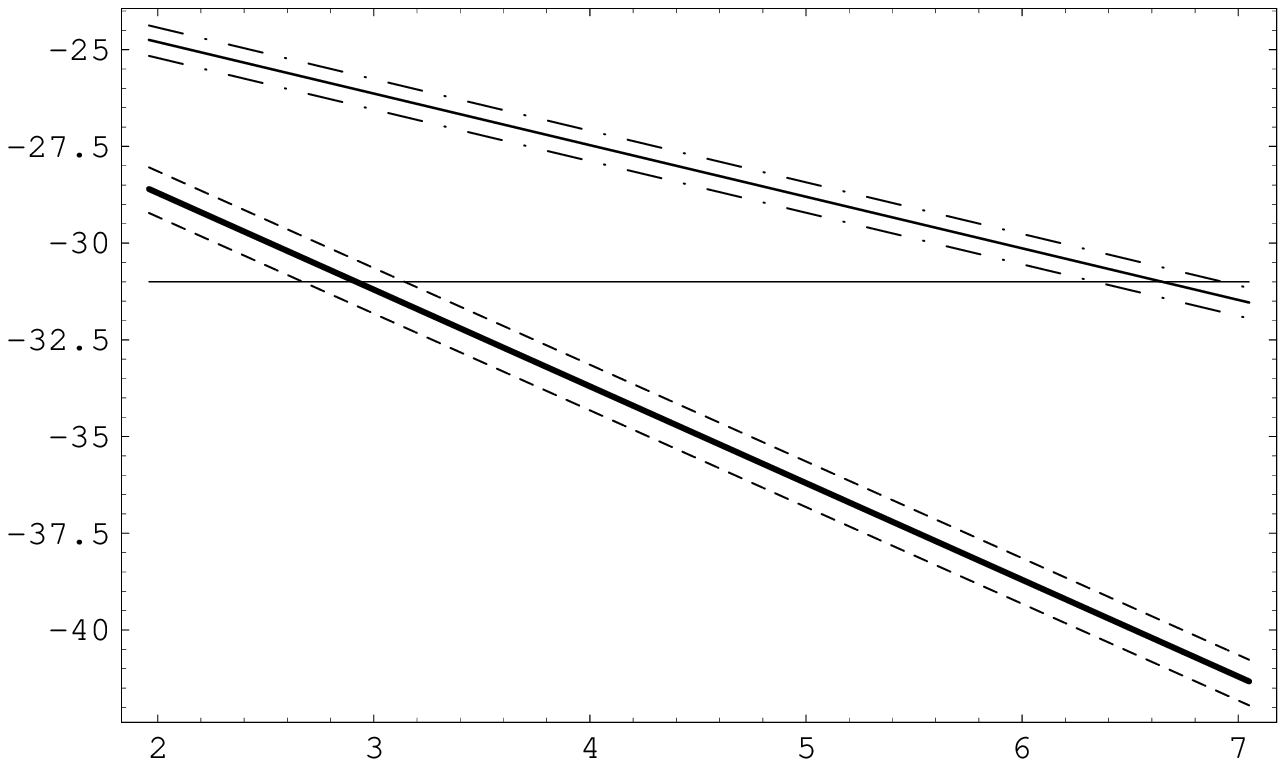}}\\
{\Large $\log[M_S]$}\\[3mm]
{\large\bfseries Fig.1}\\
\end{center}
\hspace{-1cm}{\Large $\log[\Lambda/M_{Pl}^4]$}\\
\begin{center}
{\hspace*{-20mm}\includegraphics[height=90mm,keepaspectratio=true]{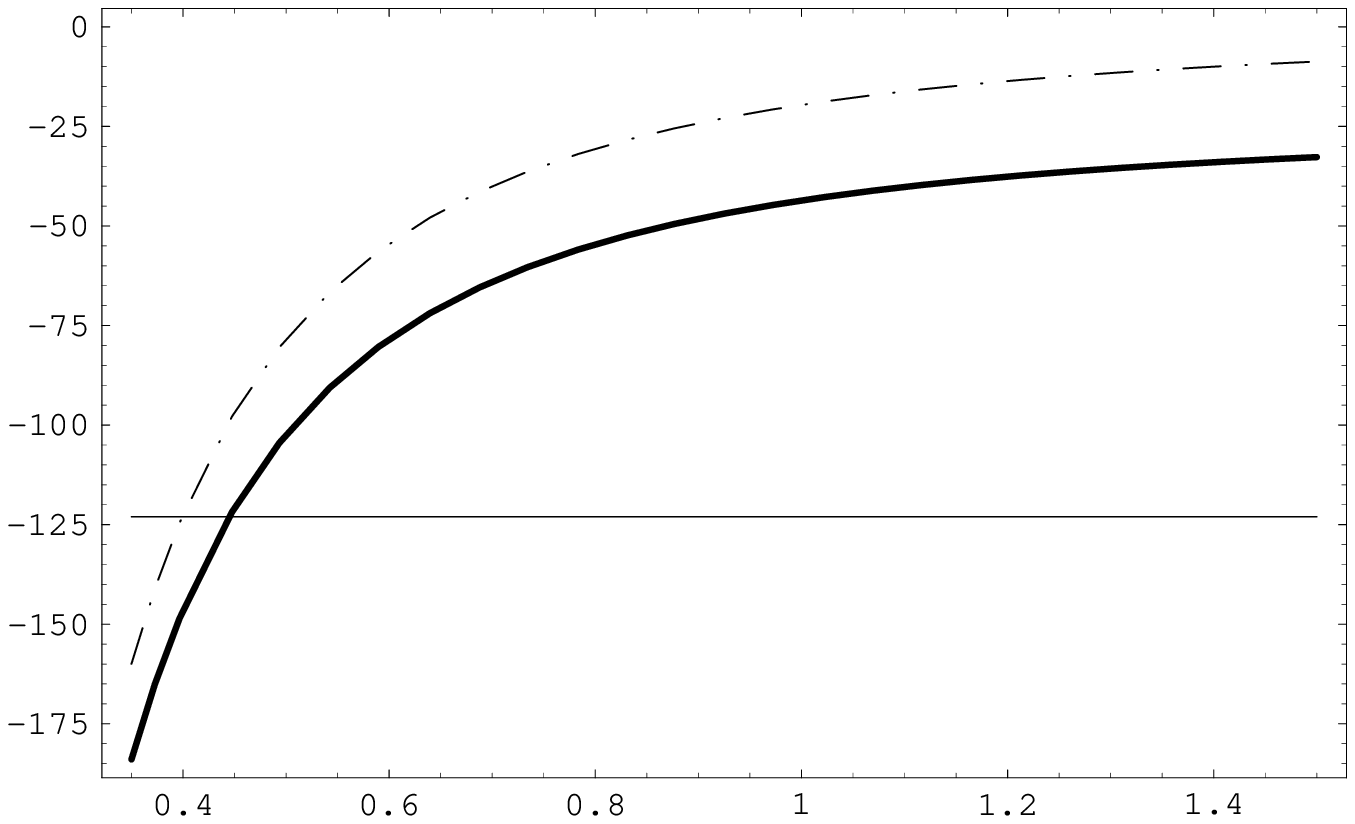}}\\
{\Large $g_{33}^{(2)}(\mu_X)$}\\[3mm]
{\large\bfseries Fig.2}
\end{center}

\newpage
\hspace{-1cm}{\Large $g_{33}(\mu_X)$}\\
\begin{center}
{\hspace*{-20mm}\includegraphics[height=90mm,keepaspectratio=true]{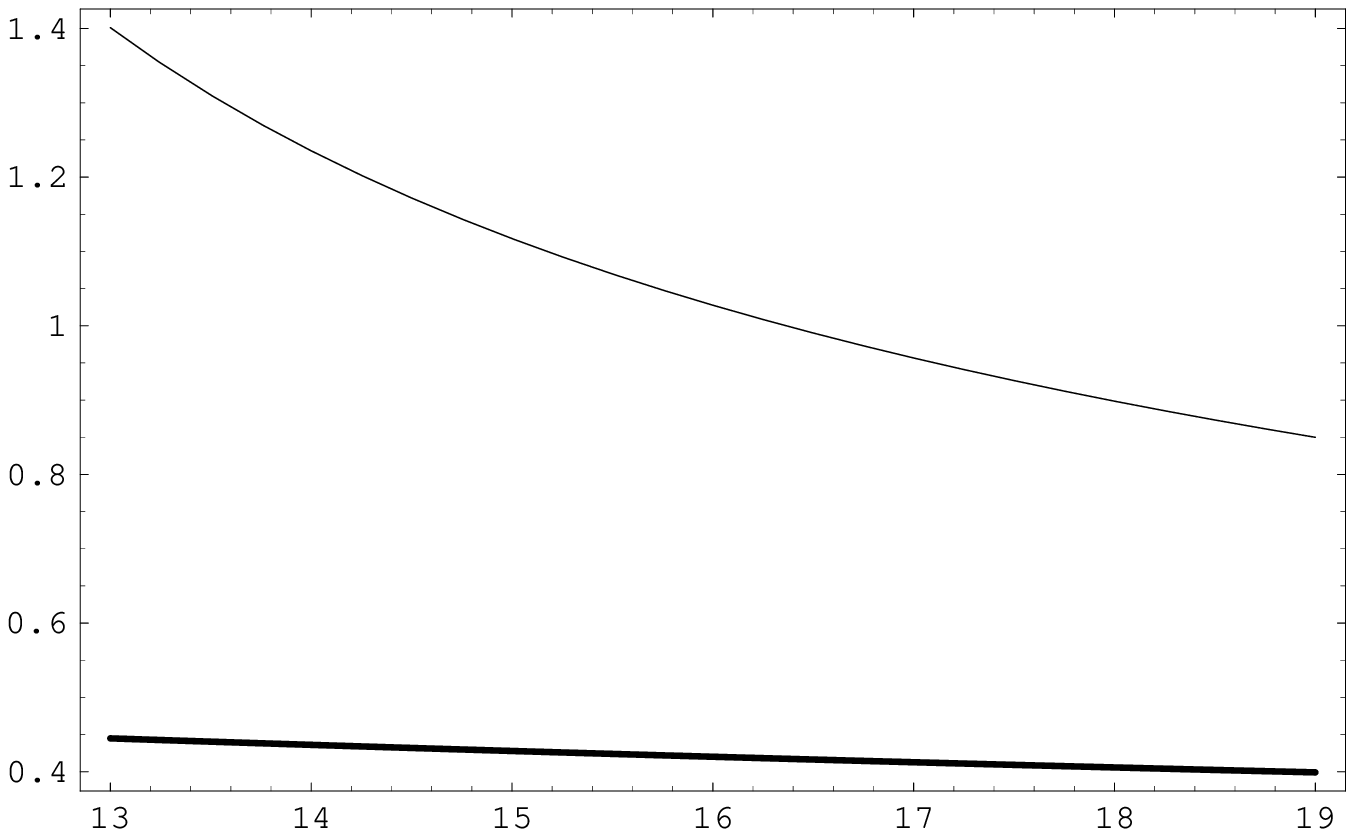}}\\
{\Large $\log[\mu_X]$}\\[3mm]
{\large\bfseries Fig.3}
\end{center}

\end{document}